\documentclass[aps,prc,twocolumn,groupedaddress,showpacs]{revtex4}

\usepackage{graphicx}
\usepackage{amssymb}
\usepackage{amsmath}
\usepackage{color}

\begin{document}


%
%
%
%
\title{Structure of $A=7-8$ nuclei with two- plus three-nucleon interactions from chiral effective field theory}


%
\author{P. Maris}
\affiliation{Department of Physics and Astronomy,
Iowa State University, Ames, Ia 50011-3160, USA}
\author{P. Navr{\'a}til}
\affiliation{TRIUMF, 4004 Wesbrook Mall, Vancouver BC, V6T 2A3, Canada}
\affiliation{Lawrence Livermore National Laboratory,
L-414, P.O. Box 808, Livermore, CA   94551, USA}
\author{J.P. Vary}
\affiliation{Department of Physics and Astronomy,
Iowa State University, Ames, Ia 50011-3160, USA}
%
%
\date{\today}
\begin{abstract}
We solve the {\it ab initio} no-core shell model (NCSM) in the complete $8\hbar\Omega$ 
($N_{\rm max}=8$) basis for $A=7$ and $A=8$ nuclei with two-nucleon and three-nucleon interactions derived within chiral effective field theory (EFT). We find that including the chiral EFT three-nucleon interaction in the Hamiltonian improves overall agreement with experimental binding energies, excitation spectra, transitions and electromagnetic moments.  We predict states
that exhibit sensitivity to including the chiral EFT three-nucleon interaction but are not yet known experimentally.
\end{abstract}

\pacs{21.30.Fe, 21.60.Cs, 27.20.+n}

\maketitle

\section{Introduction}

One of the most challenging problems in nuclear physics is to calculate nuclear properties starting from the strong interactions that accurately describe the nucleon-nucleon, three-nucleon and, possibly, four-nucleon systems. There are two major issues to overcome. First, the basic interactions among nucleons  are complicated, they are not uniquely defined and there is ample evidence that more than just two-nucleon forces are important.  Second, as a consequence of the complex nature of the inter-nucleon interactions, the quantum many-body problem for these strongly-interacting self-bound nuclei is very difficult to solve with good precision.

Interactions among nucleons are governed by QCD.
In the low-energy regime relevant to nuclear structure,
QCD is non-perturbative and hard to solve directly to obtain these inter-nucleon interactions.
New theoretical developments, however, allow us to connect QCD
with low-energy nuclear physics through the promising bridge of chiral effective field theory ($\chi$EFT)~\cite{Weinberg}.
The $\chi$EFT that includes pions but omits explicit nucleon excitations, predicts, along with the nucleon-nucleon ($NN$) interaction
at the leading order, a three-nucleon ($NNN$) interaction starting at the 3rd order 
(next-to-next-to-leading order or N$^2$LO)~\cite{Weinberg,vanKolck:1994,Epelbaum:2002}, 
and even a four-nucleon ($NNNN$) interaction starting at the 4th order (N$^3$LO)~\cite{Epelbaum06}.
The details of QCD dynamics are contained in parameters,
low-energy constants (LECs), not fixed by the symmetry. These parameters
can be constrained by experiment. A crucial feature of $\chi$EFT is the consistency between
the $NN$, $NNN$ and $NNNN$ parts. As a consequence, at N$^2$LO and N$^3$LO, except
for two LECs, assigned to two $NNN$ diagrams, the potential is fully
constrained by the parameters defining the $NN$ interaction.

We have previously performed extensive calculations for light nuclei with the $\chi$EFT $NN$ and $NNN$ interactions within the {\it ab initio} no-core shell model (NCSM)~\cite{NCSM}. In particular, we investigated $A= 6$ and $7$ nuclei~\cite{Nogga06}, $A= 6, 10, 11, 12$ and $13$ nuclei~\cite{nav}, and $A = 14$ nuclei~\cite{C14}. The major conclusion obtained from these calculations was the confirmation of the significance of the $NNN$ interaction not only for the binding energies but also for the description of excitation energies and other observables such as magnetic dipole ($M1)$ and Gamow-Teller (GT) transitions. The $NNN$ effects were found to be enhanced for the mid-$p$-shell nuclei~\cite{nav}, where, for example, the $^{10}$B ground-state spin is in agreement with experiment only when the $NNN$ interaction is included in the Hamiltonian.  One of the dramatic consequences of including $\chi$EFT  $NNN$ interactions has recently been found essential to explain the anomalous long lifetime (suppressed GT matrix element) of $^{14}$C ~\cite{C14}.  We previously discovered another dramatic consequence of $NNN$ interactions in producing a strong enhancement of the B(M1) transition from the ground state of $^{12}$C to the $(J^{\pi},T) = (1^+,1)$ excited state~\cite{Hayes03}, a transition that plays a major role in inelastic neutrino scattering.   This early demonstration of the B(M1) enhancement featured the use of two realistic $NN$ interactions, Argonne V8' ~\cite{av8p} and CD-Bonn 2000~\cite{cdb2k} each combined with the Tucson-Melbourne "prime" $NNN$ interaction ~\cite{TMprime99}.  This B(M1) enhancement has been confirmed with $\chi$EFT $NN$ and $NNN$ interactions~\cite{nav}.

These calculations were performed by employing the Okubo-Lee-Suzuki (OLS) effective interaction approach~\cite{Okubo,LS80} primarily in the $6\hbar\Omega$ ($N_{\rm max}=6$) basis space.  The exceptions are the $A=6$ and $A=14$ results which were obtained in the $8\hbar\Omega$ space. It is desirable to extend all the calculations to larger basis sizes for several reasons. First, one would like to check the convergence of the smaller-space calculations. Second, the soft similarity-renormalization-group (SRG) evolved interactions are now available including the $NNN$ terms~\cite{SRG_3b,Li6_SRG}. Variational calculations with these interactions require bases bigger than just $6\hbar\Omega$ to fully establish the systematic trends. 
Also, the trends in results from different renormalization schemes, such as OLS and SRG, need to be compared with each other to better understand their advantages and drawbacks.
Third, the importance-truncation approach has been successfully implemented for the NCSM calculations~\cite{IT-NCSM,Roth09}.  That approach requires benchmarking 
against exact calculations in the same $N_{\rm max}\hbar\Omega$ basis space.
Fourth, the NCSM has been extended by the resonating group method (NCSM/RGM) for the description of nuclear reactions~\cite{NCSMRGM}. The NCSM/RGM approach relies on the SRG interactions and requires basis expansion beyond $6\hbar\Omega$. 

We report here our recent technical advances that allow us to perform the full-space $8\hbar\Omega$ calculations for all the $p$-shell nuclei. These technical advances include both the developments of the configuration interaction code MFDn~\cite{MFDn} and the development of the codes that calculate the $NNN$ matrix elements. In this paper, we present results for $A=7,8$ nuclei using the OLS method. Calculations for other $p$-shell nuclei using both the OLS and the SRG methods are underway and will be reported separately.

In Sect.~\ref{NCSM}, we briefly describe the NCSM approach and the technical advances that allowed us to extend the basis size of the calculations. Results for $A=7$ and $A=8$ nuclei are given in Sect.~\ref{A_7_8}. Conclusions are drawn in Sect.~\ref{conclusions}.

\section{{\it Ab initio} no-core shell model}
\label{NCSM}

In the {\it ab initio} NCSM, we consider a system 
of $A$ point-like non-relativistic nucleons that interact by realistic 
$NN$ or $NN+NNN$ interactions. Unlike in standard shell model calculations,
in the NCSM there is no inert core, all the nucleons are considered active -
therefore the ``no-core'' in the name of the approach.
Besides the employment of realistic $NN$ or $NN+NNN$ interactions, two other major features 
characterize the NCSM: i) the use of an harmonic oscillator (HO) basis 
truncated by a chosen maximal HO excitation energy $N_{\rm max}\hbar\Omega$ 
(equivalently, the number of HO quanta $N_{\rm max}$) above
the unperturbed ground state (i.e. the lowest possible HO configuration)
of the $A$-nucleon system; 
and ii) the use of effective interactions.
The reason behind the choice of the HO basis is the fact that
this is the only basis known (aside from the plane wave basis) that allows one to use single-nucleon coordinates and consequently 
the second-quantization representation without violating the translational invariance
of the system. The powerful techniques based on second quantization and developed 
for standard shell model calculations can then be utilized - therefore the ``shell model''
in the name of the approach. As a downside, one has to face the consequences of the incorrect
asymptotic behavior of the HO basis. 
The preservation of translational invariance is a consequence of the $N_{\rm max}\hbar\Omega$ truncation.

In order obtain a reasonable approximation in a finite basis space (characterized by $N_{\rm max}$ and $\hbar\Omega$) to the exact results in a complete (but infinite-dimensional) basis space,
we construct an OLS effective interaction
from the original realistic $NN$ or $NN+NNN$ potentials by means of a unitary transformation.
We carry out this transformation at the two-body level (``NN only") and at the level including both NN and NNN interactions (``NN + NNN").  In principle, one can perform the unitary transformation and generate many-body interactions up to and including all nucleons.   However, going beyond NN + NNN is technically very challenging.  We may refer to our implementation of OLS as the cluster-truncated OLS approach.

The OLS effective interaction depends on the basis parameters ($N_{\rm max}$,$\Omega$) and recovers
the original realistic $NN$ or $NN+NNN$ interaction as $N_{\rm max}$ approaches infinity.
In principle, one can also perform calculations with the unmodified, ``bare'', original 
interactions. Such calculations are then variational with respect to $N_{\rm max}$ and $\Omega$.

In this work, we use $NN$ and $NNN$ interactions derived within the chiral EFT. In particular, we employ the chiral N$^3$LO $NN$ interactions from Ref.~\cite{Entem:2003ft,Machleidt:2011zz} and the chiral N$^2$LO $NNN$ interaction~\cite{Epelbaum:2002} in the local form of Ref.~\cite{NavratilFB07}. For the low-energy constants of the $NNN$ interaction not fixed by the two-nucleon data, we adopt values that reproduce the triton binding energy and half life~\cite{Gazit:2008ma}. Next, we calculate three-body effective interaction from the chiral $NN+NNN$ interactions using the OLS procedure. As mentioned above, we adopt a cluster truncation which means that the three-body interaction is derived from full-space three-nucleon system solutions and the resulting three-body effective interaction is then input into the shell model code for the $A$-nucleon system.  A large-scale diagonalization is then performed in the $A$-nucleon $N_{\rm max}\hbar\Omega$ HO basis.

As the three-body effective interactions are derived in the Jacobi-coordinate HO basis but the $p$-shell nuclei calculations are performed using the shell model code in a Cartesian-coordinate single-particle Slater-determinant $M$-scheme basis, we need to perform a suitable 
transformation of the interactions. This transformation is a generalization of the well-known transformation on the two-body level that depends on HO Brody-Moshinsky brackets. Details of this transformations are given in Refs.~\cite{v3eff,v3beff,Nogga06}. In this work, we use the particular version given in the appendix of Ref.~\cite{Nogga06}. The corresponding computer code was improved compared to earlier applications~\cite{nav} that allow us now to perform the transformations up to $N_{\rm max}=8$ basis spaces for all $p$-shell nuclei. We note that for the $p$-shell nuclei with $A\ge 7$, the number of $NNN$ $M$-scheme matrix elements increases compared to the $A=6$ case that was handled up to $N_{\rm max}=8$ already in Ref.~\cite{nav}.  We note that the Jacobi-to-Slater-determinant $NNN$ transformation was further improved recently by utilizing a factorization with a $NNN$ coupled-JT scheme~\cite{Roth} allowing to reach still larger $N_{\rm max}$ spaces.

It is a challenge to utilize the M-scheme $NNN$ interaction in a shell model code. First, one has to deal with a large number of the $NNN$ matrix elements and, second, the number of the non-zero many-body Hamiltonian matrix elements increases by more than an order of magnitude compared to calculations with just $NN$ interactions. Both these issues were successfully addressed in the newer versions of the shell model code MFDn, a hybrid OpenMP and MPI code \cite{MFDn} .  The calculations discussed in this paper were performed on Franklin at NERSC using up to six thousand cores for the largest runs.  Current versions of MFDn are capable of handling dimensions exceeding 1 billion with $NN+NNN$ Hamiltonians. This is sufficient for the $N_{\rm max}=8$ basis space of any $p$-shell nucleus.  The largest calculations to date are the $N_{\rm max}=8$ results for $^{14}$C and $^{14}$N presented in Ref.~\cite{C14}.

\section{Applications to $A=7$ and $A=8$ nuclei}
\label{A_7_8}

Our calculations for both $A=7$ and $A=8$ nuclei were performed in model spaces
up to $8\hbar\Omega$ for a wide range of HO frequencies. We then selected the HO frequency that corresponds to the ground-state energy minimum in the $8\hbar\Omega$ space for detailed comparison of our results with experimental data.
To further elucidate convergence properties,  we discuss the dependencies of some observables on the basis space parameters ($N_{max}$,$\Omega$). 

\begin{figure}[t]
\includegraphics*[width=1.1\columnwidth]{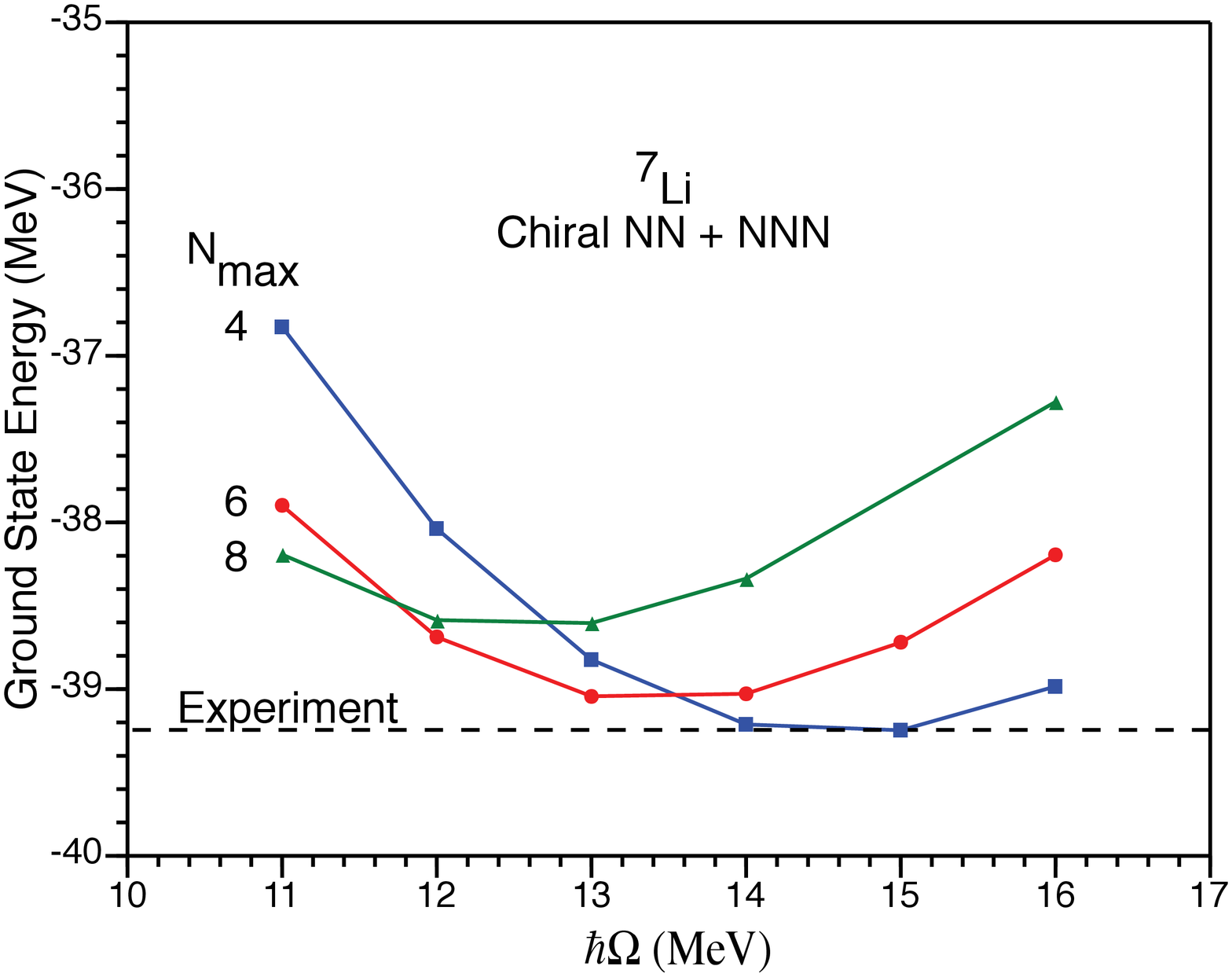}
\caption{(Color online) Calculated ground state energy of $^{7}$Li in the NCSM with 
chiral EFT $NN$ and $NNN$ interactions that reproduce the triton binding energy and half life. 
The dependence on the HO frequency and size of the basis is presented.}
\label{7Li_NN_NNN_gs}
\end{figure}
\begin{figure}[t]
\includegraphics*[width=1.1\columnwidth]{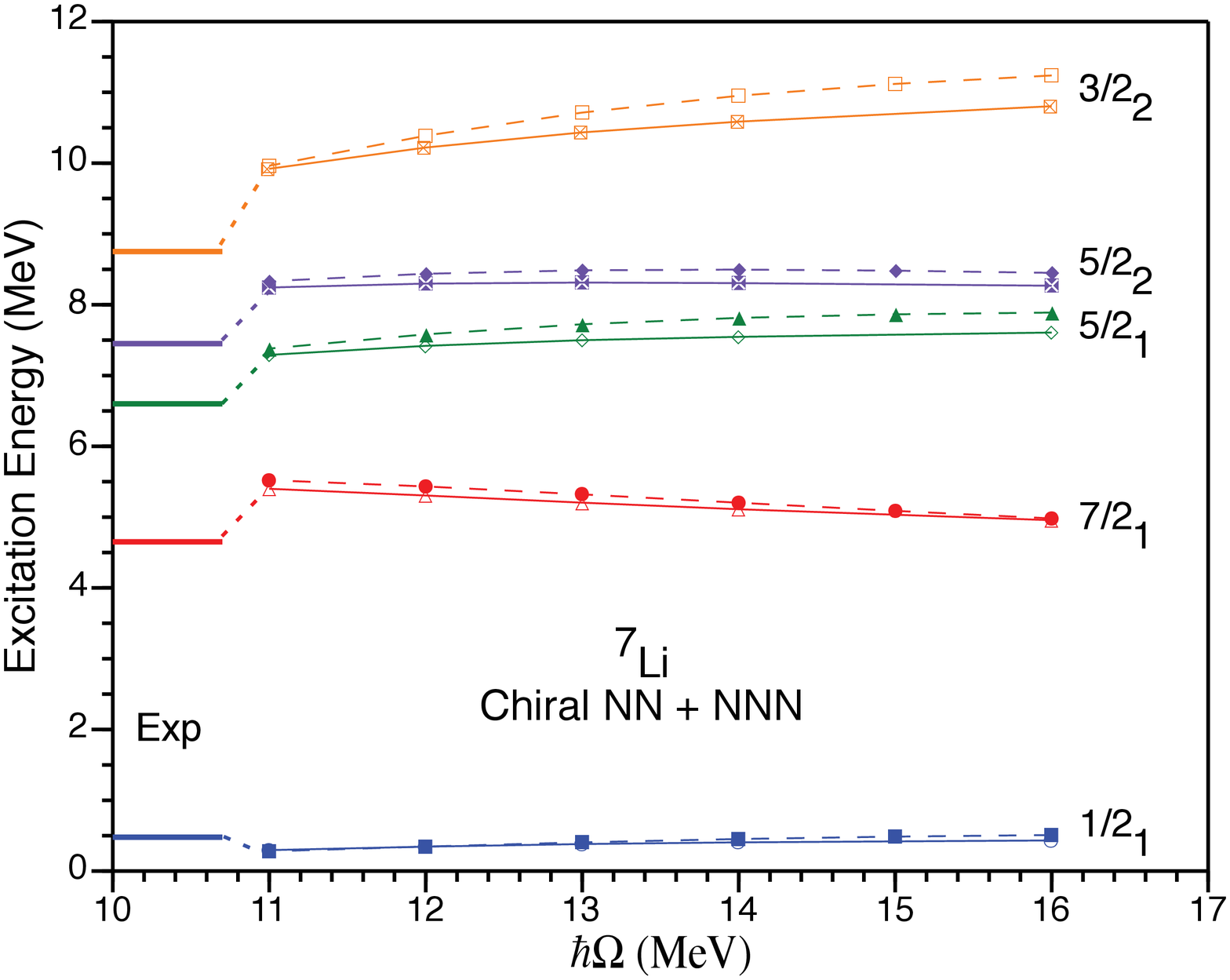}
\caption{(Color online) Calculated excitation energies of the lowest 5 excited states of $^{7}$Li
in the NCSM with chiral EFT $NN$ and $NNN$ interactions that reproduce the triton binding energy 
and half life. The dependence on the HO frequency is presented at the two highest values of basis space truncation $N_{\rm max}=6$ (dashed lines) and $8$ (solid lines).
The experimental excitation energies are shown on the left for comparison.}
\label{7Li_Chiral_OLS_Convergence}
\end{figure}

The $^7$Li ground-state energy dependence on the HO frequency for different model spaces
is shown in Fig. \ref{7Li_NN_NNN_gs} for the 
chiral $NN+NNN$ interaction.
We observe that the minimum shifts towards smaller HO frequency as the basis space increases. 
There is also a weak but irregular trend towards less dependence on the HO frequency as would be expected from reduced roles of the neglected 4-body and higher-body interactions.  To be specific, 
Fig. \ref{7Li_NN_NNN_gs}
shows that the ground state (gs) energy covers a range of $(2.42, 1.14, 1.33)$ MeV for $N_{\rm max}=(4,6,8)$ in the range of $\hbar\Omega$ depicted.

The pattern of the $N_{\rm max}=(4,6)$ curves in Fig. \ref{7Li_NN_NNN_gs} is very similar to those in 
Ref.~\cite{Nogga06} where different parameters for the chiral NNN interaction were employed for $^7$Li.  There is a  shift with the present interaction of about 1 MeV at $N_{\rm max}=6$ towards greater binding and towards better agreement with experiment when compared with the results of Ref.~\cite{Nogga06} with the chiral NNN interaction closest (called "3NF-A" in Ref.~\cite{Nogga06} ) to the present case.

We show in Fig. \ref{7Li_Chiral_OLS_Convergence} the dependence of the low-lying excited states of  $^7$Li on the harmonic oscillator energy at the two highest values of basis space truncation, $N_{\rm max}=6$ and $8$.   Fig. \ref{7Li_Chiral_OLS_Convergence} demonstrates the systematic trend to improved independence of $\hbar\Omega$ with increasing $N_{\rm max}$ for the excitation spectra (slopes of the excitation energies decrease with increasing $N_{\rm max}$).  Furthermore, the shifts in the excitation energies when proceeding from $N_{\rm max}=6$ to $8$ are less than the spread in the $\hbar\Omega$ dependence at $N_{\rm max}=6$ over the range of $\hbar\Omega$ depicted in Fig. \ref{7Li_Chiral_OLS_Convergence}.  

Quantifying the uncertainties in our results for nuclear observables is a major challenge.  The systematic uncertainties due to lack of  complete convergence dominate our overall uncertainties by at least an order of magnitude. The actual uncertainties are dependent on the specific observable as well as whether the observable is hindered or enhanced compared to phenomenological single-particle values.  To give an example, we estimate the numerical uncertainties in our calculated gs energies and excitation energies to be around $(50,1)$ keV arising respectively from the numerical evaluation of (1) the effective NN or NN+NNN interactions for the selected basis space and (2) the numerical solution of the many-body eigenvalue problem. However, as we shall discuss now, our overall uncertainties are dominated by incomplete convergence with increasing basis size.  Since we do not have converged calculations to calibrate these uncertainties, we will simply quantify specific measures of our basis space dependence.  Convergence arises when these measures of basis space dependence vanish.

The results in Fig. \ref{7Li_Chiral_OLS_Convergence} provide indications of our basis space dependence.  For the present work, we adopt the following procedure to estimate the dependence of the excitation energies on the basis-space truncation.  We quote two quantities: (1) one half of the total spread in the excitation energy over the range in $\hbar\Omega$ shown in this figure at $N_{\rm max}=8$, and (2) the total shift in excitation energy obtained from the increment of $N_{\rm max}=6$ to $8$ at the selected optimum frequency. These quantities are quoted in parenthesis beside each excitation energy result for the $NN + NNN$ interaction in the tables below.  For the excited states in the present work
we present both estimates in the above respective order to show the state-by-state fluctuations in each quantity.  For uniformity and completeness, we present two significant figures for each estimate. 

Since many of the states we investigate are resonances and our basis lacks explicit coupling
to the continuum, we expect, in accordance with the findings of Ref. \cite{Shirokov09}, that broader resonances are associated with larger $\hbar\Omega$-dependence in the HO basis
calculations.  Thus, one may also interpret our first measure of basis-space dependence as a rough
indicator of the resonance width.  This may be useful for estimating relative widths~\cite{Shirokov09}.

From the $N_{\rm max}=8$ curve in Fig. \ref{7Li_NN_NNN_gs} we select the optimal frequency as $\hbar\Omega=13$ MeV for examining our results in greater detail. 
This adoption sets one of the inputs to the determination of the basis-space dependence in excitation energies as just described.  We also define the basis-space dependence of our total gs energy as the difference in total energy at this adopted minimum for the basis space increment from $N_{\rm max}=6$ to $8$. As an example, this produces the estimate of $0.44$ MeV for the $^7$Li gs energy which is quoted in parenthesis next to the eigenvalue in Table ~\ref{tab:Be7energies}. 

We observe a similarity in the $N_{\rm max}$ dependence or results in Figs. \ref{7Li_NN_NNN_gs}
and \ref{7Li_Chiral_OLS_Convergence}.  In both cases, our estimated uncertainties range up to several hundred keV (see Table~\ref{tab:Be7energies}).  However, in the absence of a firm trend in $N_{\rm max}$ for our results, one should not take our quoted uncertainties as estimates of numerical accuracy but rather as characteristics of the dependence of the results on the presently available basis spaces.

\begin{figure}[t]
\includegraphics*[width=1.0\columnwidth]{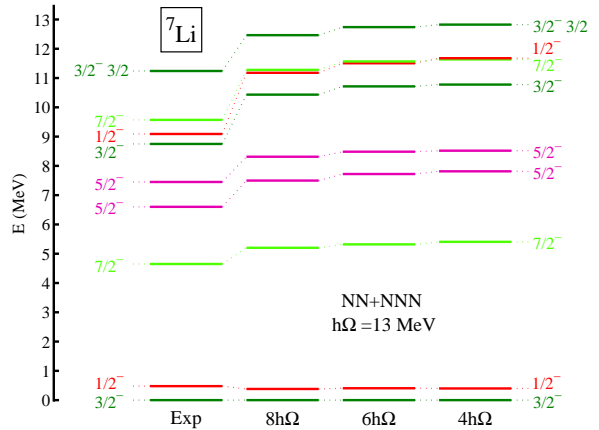}
\caption{(Color online) Calculated and experimental excitation energies of $^{7}$Li. Dependence on the size of the basis is presented. The chiral EFT $NN$ and $NNN$ interaction was used. The isospin of the states is $T=1/2$ unless shown otherwise. See the text for further details.}\label{fig:Li7_spectrum}
\end{figure}

We show the low-lying spectra of $^7$Li in Fig. \ref{fig:Li7_spectrum} at the optimum frequency and at the sequence of $N_{\rm max}$ truncations corresponding to the curves in Fig.  \ref{7Li_NN_NNN_gs}.
The energies, radii and electromagnetic observables are summarized
in Tables~\ref{tab:Be7energies} and \ref{tab:Be7moments}, where we
also include the $^7$Be results. We obtain the same level ordering
for $^7$Be and $^7$Li which is also the same for both $NN$ and the $NN+NNN$ interactions
with the exception of a reversal of the $7/2^-_2$ and $3/2^-_2$ levels in $^7$Be. 
That is, in $^7$Be, the experimental $7/2^-_2$ and
$3/2^-_2$ levels are reversed compared to our results and the
situation in $^7$Li.
On the other hand, our $NN+NNN$ ordering is in agreement with experiment for the
9 lowest states in $^7$Li. 

Our calculated spectra for both of the $A=7$ nuclei show 
a reasonable stability with respect to the frequency change.  
The results in Table~\ref{tab:Be7energies} (and $A=8$ results in Tables~\ref{tab:B8Li8energies} and \ref{tab:Be8energies} below) indicate that
there are residual differences between theoretical and experimental energies that are significantly larger than our quoted basis-space dependence of the calculated results.   
It will be interesting to see if the differences between theory and experiment 
persist once more accurate calculations become feasible.  
If they do, the question becomes whether these differences 
are significantly reduced, for example, when a chiral NNN interaction becomes available that is more complete than the one currently available~\cite{Epelbaum_priv_comm}.

\begin{table}[hbtp]
  \caption{The $^7$Be and $^7$Li ground and excited state energies (in MeV)
obtained using the chiral $NN$ and chiral $NN+NNN$ interactions.
The HO frequency of $\hbar\Omega=13$ MeV and the $8\hbar\Omega$ 
model space were used.  Our measures of basis-space dependence are given
for the last two significant figures of the quoted theory result. 
Two quantities, as explained in the text, are quoted in parenthesis
for excitation energies with the notation: (0.5 x total 
range of swing with $\hbar\Omega$ at $N_{max}=8$; 
difference at $\hbar\Omega = 13$ MeV 
between $N_{max}=6$ and $8$ results). Only the second quantity is
quoted for the magnitude of the total ground state energy.  
The $^7$Be states labelled ``mixed iso" have large isospin mixing 
and their basis-space dependence can be approximated by the dependences in the 
corresponding states of $^7$Li.
Experimental values are from Ref. \cite{A=5-7}. 
  \label{tab:Be7energies}}
  \begin{ruledtabular}
    \begin{tabular}{cccc}
\multicolumn{4}{c}{$^7$Be} \\
                                        & Expt.     &  $NN$   & $NN+NNN$ \\
                                         \hline
$|E_{\rm gs}(\frac{3}{2}^- \frac{1}{2})|$ & 37.6004(5) & 32.75 & 36.98(43) \\
$E_{\rm x}(\frac{1}{2}^-_1 \frac{1}{2})$ & 0.429       &  0.233   &  0.371 (67;24) \\
$E_{\rm x}(\frac{7}{2}^-_1 \frac{1}{2})$ & 4.57(5)    &  5.28   &  5.14 (21;11) \\
$E_{\rm x}(\frac{5}{2}^-_1 \frac{1}{2})$ & 6.73(10)  &  6.66   &  7.43 (17;23) \\
$E_{\rm x}(\frac{5}{2}^-_2 \frac{1}{2})$ & 7.21(6)    &  8.12   &  8.11 (04;18) \\
$E_{\rm x}(\frac{7}{2}^-_2 \frac{1}{2})$ & 9.27(10)  & 10.52 & 10.98 (25;31) \\
$E_{\rm x}(\frac{3}{2}^-_2 \frac{1}{2})$ & 9.9           &  9.29   & 10.13 (46;30) \\
$E_{\rm x}(\frac{1}{2}^-_2 \frac{1}{2})$ &                  & 10.00  & 10.91 (49;35) \\
$E_{\rm x}(\frac{3}{2}^-_3 \frac{1}{2})$ &                  & 11.57  &  12.28 (mixed iso)\\
$E_{\rm x}(\frac{3}{2}^-_1 \frac{3}{2})$ & 11.01(3)  & 12.10  &  12.38 (mixed iso)\\
\hline
\multicolumn{4}{c}{$^7$Li} \\
                                        & Expt.   & $NN$ & $NN+NNN$ \\ 
                                        \hline
$|E_{\rm gs}(\frac{3}{2}^- \frac{1}{2})|$ & 39.245  & 34.34 & 38.60(44) \\
$E_{\rm x}(\frac{1}{2}^-_1 \frac{1}{2})$ & 0.478    &  0.238 &  0.382  (69;24) \\
$E_{\rm x}(\frac{7}{2}^-_1 \frac{1}{2})$ & 4.65      &  5.36 &  5.20  (22;12) \\
$E_{\rm x}(\frac{5}{2}^-_1 \frac{1}{2})$ & 6.60      &  6.72 &  7.50  (16;23) \\
$E_{\rm x}(\frac{5}{2}^-_2 \frac{1}{2})$ & 7.45      &  8.35 &  8.31  (01;17) \\
$E_{\rm x}(\frac{3}{2}^-_2 \frac{1}{2})$ & 8.75      &  9.58 & 10.43  (44;28) \\
$E_{\rm x}(\frac{1}{2}^-_2 \frac{1}{2})$ & 9.09      & 10.29 & 11.18  (47;33) \\
$E_{\rm x}(\frac{7}{2}^-_2 \frac{1}{2})$ & 9.57      & 10.81 & 11.28  (24;29) \\
$E_{\rm x}(\frac{3}{2}^-_1 \frac{3}{2})$ & 11.24    & 12.25 & 12.46  (18;28) \\
    \end{tabular}
  \end{ruledtabular}
\end{table}

We present in Table  \ref{tab:Be7moments} a selection of results for magnetic moments, 
M1 transitions and other properties of the $A=7$ nuclei.  All electromagnetic observables
are evaluated with the free-space electromagnetic coupling constants.  
That is, we do not employ effective
charges or effective magnetic moments for the nucleons. 

The results in Table  \ref{tab:Be7moments} 
with $NN$ alone and $NN + NNN$ 
interactions are both in reasonable agreement with experiment. 
One observes that there is a trend for radii and quadrupole moments
to increase with increasing basis size and/or decreasing frequency.
This is, in part, a consequence of the incorrect asymptotics of the HO basis
and also our basis space truncation. 
Convergence rates for the radii and quadrupole moment appear improved 
at a smaller HO frequency since they are less dependent on the basis truncation.

We have performed various tests to establish that our calculated electroweak observables,
those near or greater than single-particle values, are accurate to three significant
digits. However, we are not able to quantify the basis-space dependence for our electroweak observables at the present time.  By presenting results at different values of $\hbar\Omega$ and $N_{max}$ we provide a preliminary indication of those dependences. Quantifying their systematic uncertainties more rigorously will require an extensive separate investigation. In the meantime, our best estimate of the exact theoretical result is the one obtained in the largest basis space with the optimum value of $\hbar\Omega$.  

The magnetic moments and B(M1) values tend to be about $10\%$ to $20\%$ smaller 
in magnitude than experiment.  We believe this is an acceptable range of difference 
since we have not included exchange current corrections which can easily be in
the range needed to improve the agreement with experiment.   Future work will
address these corrections.  In the meantime, we can offer support for this
belief by citing the corrections due to two-body currents obtained 
in the {\it ab initio} evaluation of the the magnetic moment of the $^7$Li gs
using GFMC techniques with AV18 plus Illinois-2 three-body forces ~\cite{Marcucci:2008mg}. In that
investigation, the two-body currents raised the 
$^7$Li gs magnetic moment from 2.9 $\mu_N$ to 3.2 $\mu_N$.
Similarly, the two-body currents changed the $^7$Be gs magnetic moment
from -1.06 $\mu_N$ to -1.49 $\mu_N$.  Both these corrections are in the direction
and of the magnitude needed to explain the difference of our results from experiment.

We adopt the experimental gs rms charge radius for $^7$Li (2.44(4) fm) 
and $^8$Li (2.34(5) fm)
from a recent detailed analysis of the experimental and theoretical 
gs properties of the Lithium isotopes \cite{Nortershauser:2011zz}.
We also adopt the corrections they define and evaluate (finite proton charge density,
neutron charge density, etc.) in order to extract a point proton rms radius $r_{\rm p}$ from
the measured rms charge radius that we quote in our tables
as the experimental value for $r_{\rm p}$ to be compared with our theoretical results.
Note that our theoretical $r_{\rm p}$ is free of spurious cm motion effects.

%
%
%
%
%
%
\begin{table}[hbtp]
  \caption{The $^7$Be and $^7$Li point-proton rms radii (in fm), ground-state quadrupole (in $e$fm$^2$),
magnetic moments (in $\mu_{\rm N}$) and M1 transitions (in $\mu_{\rm N}^2$)
obtained within the NCSM for different HO frequencies (given in MeV)
and model spaces for the chiral $NN$ and chiral $NN+NNN$ interactions. 
Most experimental values are from Ref. \cite{A=5-7}.
The point proton rms
radius $r_{\rm p}$ for $^7$Be is evaluated from the experimental rms charge radius
of 2.647(17) fm from Ref \cite{Nortershauser:2008vp} using corrections of Ref \cite{Nortershauser:2011zz} as discussed in the text.
The point proton rms
radius $r_{\rm p}$ for $^7$Li is evaluated from the experimental rms charge radius
of Ref \cite{Nortershauser:2011zz} as discussed in the text.
  \label{tab:Be7moments}}
  \begin{ruledtabular}
    \begin{tabular}{cccccc}
\multicolumn{6}{c}{$^7$Be} \\ \hline
\multicolumn{6}{c}{chiral $NN$} \\
$\hbar\Omega$ & $N_{\rm max}$ & $r_{\rm p}$ & Q & $\mu$ & B(M1; $\frac{1}{2}^-\rightarrow
\frac{3}{2}^-$)\\
\hline
13 &  4 & 2.281 & -4.484 & -1.157 &  3.196 \\
13 &  6 & 2.301 & -4.798 & -1.147 &  3.142 \\
13 &  8 & 2.345 & -5.125 & -1.138 &  3.094 \\ \hline
\multicolumn{2}{c}{Expt.} &  2.52(3) & - - -  & -1.398(15) & 3.71(48) \\ \hline
\multicolumn{6}{c}{chiral $NN+NNN$} \\
$\hbar\Omega$ & $N_{\rm max}$ & $r_{\rm p}$ & Q & $\mu$ & B(M1;$\frac{1}{2}^-\rightarrow
\frac{3}{2}^-$)\\
\hline
11 &  4 & 2.379 & -4.701 & -1.151 & 3.250 \\
11 &  6 & 2.351 & -4.816 & -1.146 & 3.196 \\
11 &  8 & 2.355 & -4.982 & -1.137 & 3.144 \\ \hline
%
13 &  4 & 2.237 & -4.131 & -1.153 & 3.220 \\
13 &  6 & 2.242 & -4.350 & -1.139 & 3.160 \\
13 &  8 & 2.276 & -4.615 & -1.127 & 3.106 \\ \hline
15 &  4 & 2.143 & -3.760 & -1.146 & 3.186 \\
15 &  6 & 2.180 & -4.073 & -1.129 & 3.128 \\
\hline
\multicolumn{2}{c}{Expt.} & 2.52(3) & - - -  & -1.398(15) & 3.71(48) \\
\hline
\multicolumn{6}{c}{$^7$Li} \\ \hline
\multicolumn{6}{c}{chiral $NN$} \\
$\hbar\Omega$ & $N_{\rm max}$ & $r_{\rm p}$ & Q & $\mu$ & B(M1; $\frac{1}{2}^-\rightarrow
\frac{3}{2}^-$)\\
\hline
13 &  4 & 2.130 & -2.563 &  3.038 & 4.268 \\
13 &  6 & 2.140 & -2.786 &  3.019 & 4.178 \\
13 &  8 & 2.176 & -2.987 &  3.003 & 4.100 \\ \hline
\multicolumn{2}{c}{Expt.} & 2.32(5) & -4.06(8) & +3.256 & 4.92(25) \\ \hline
\multicolumn{6}{c}{chiral $NN+NNN$} \\
$\hbar\Omega$ & $N_{\rm max}$ & $r_{\rm p}$ & Q & $\mu$ & B(M1; $\frac{1}{2}^-\rightarrow
\frac{3}{2}^-$)\\
\hline
11 &  4 & 2.225 & -2.683 &  3.035 & 4.324 \\
11 &  6 & 2.189 & -2.811 &  3.023 & 4.230 \\
11 &  8 & 2.187 & -2.936 &  3.009 & 4.144 \\ \hline
%
13 &  4 & 2.091 & -2.422 &  3.034 & 4.260 \\
13 &  6 & 2.086 & -2.587 &  3.012 & 4.154 \\
13 &  8 & 2.114 & -2.752 &  2.993 & 4.068 \\ \hline
15 &  4 & 2.002 & -2.252 &  3.028 & 4.188 \\
15 &  6 & 2.030 & -2.449 &  2.995 & 4.088 \\
\hline
\multicolumn{2}{c}{Expt.} & 2.32(5) & -4.06(8) & +3.256 & 4.92(25) \\
    \end{tabular}
  \end{ruledtabular}
\end{table}

\begin{figure}[t]
\includegraphics*[width=1.0\columnwidth]{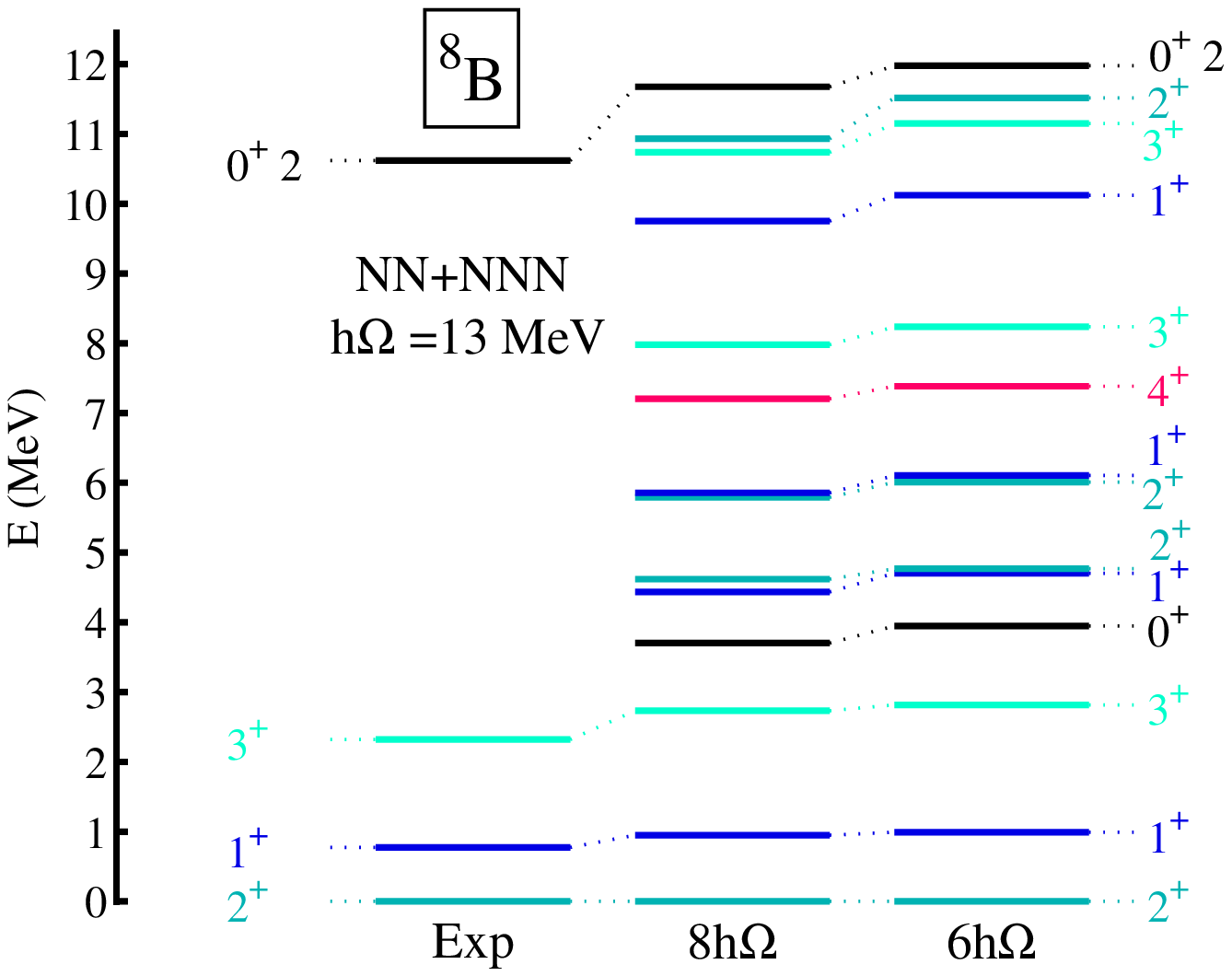}
\caption{(Color online) Calculated and experimental excitation energies of $^{8}$B. Dependence on the size of the basis is presented. The chiral EFT $NN$ and $NNN$ interaction was used. The isospin of the states is $T=1$ unless shown otherwise. See the text for further details.}\label{B8_NN_NNN_13}
\end{figure}

We present our $^8$B and $^8$Li ground-state and excited-state energy results 
in Table \ref{tab:B8Li8energies}.
The basis size dependence of the $^8$B spectra calculated using the 
chiral $NN+NNN$ interaction and the optimal HO frequency of $\hbar\Omega=13$ MeV
is shown in Fig. \ref{B8_NN_NNN_13}. 
Similar conclusions can be drawn as for the $A=7$ nuclei concerning convergence.
The dependence of the gs energy on the basis size and the HO frequency
is somewhat larger than was observed for the $A=7$ nuclei.  This may be due, in part,
to greater proximity to breakup thresholds in the $A=8$ nuclei we investigate here.
\begin{table}[hbtp]
  \caption{The $^8$B and $^8$Li ground and excited state energies (in MeV)
obtained using the chiral $NN$ and chiral $NN+NNN$ interactions.
The HO frequency of $\hbar\Omega=13$ MeV and the $8\hbar\Omega$ 
model space were used.  See the caption to table \ref{tab:Be7energies} 
and the text for explanation of the basis-space dependences quoted in
parenthesis. Experimental values are from Ref. \cite{A=8}.
  \label{tab:B8Li8energies}}
  \begin{ruledtabular}
    \begin{tabular}{cccc}
\multicolumn{4}{c}{$^8$B} \\
                                  & Expt.   & $NN$ & $NN+NNN$ \\ 
                                  \hline
$|E_{\rm gs}(2^+ 1)|$    & 37.7378(11) & 31.38 & 36.35 (67) \\
$E_{\rm x}(1^+_1 1)$       &  0.7695(25) &  0.81 &  0.95 (16;04) \\
$E_{\rm x}(3^+_1 1)$       &  2.32(20) &  2.83 &  2.73 (15;09) \\
$E_{\rm x}(0^+_1 1)$       &                  &  2.29  &  3.70 (80;25) \\
$E_{\rm x}(1^+_2 1)$       &                  &  3.11  &  4.44 (82;27) \\
$E_{\rm x}(2^+_2 1)$       &                  &  3.66  &  4.62 (44;15) \\
$E_{\rm x}(2^+_3 1)$       &                  &  5.11  &   5.79 (33;22) \\
$E_{\rm x}(1^+_3 1)$       &                  &  4.64  &  5.85 (66;25) \\
$E_{\rm x}(4^+_1 1)$       &                  &  6.27  &  7.20 (34;18) \\
$E_{\rm x}(3^+_2 1)$       &                  &  7.12  &  7.98 (47;26) \\
$E_{\rm x}(0^+_1 2)$       & 10.619(9) & 11.15 & 11.68 (27;30) \\
\hline
\multicolumn{4}{c}{$^8$Li} \\
                                  & Expt.   & $NN$ & $NN+NNN$ \\ 
                                  \hline
$|E_{\rm gs}(2^+ 1)|$      &  41.277   & 34.86 & 39.95 (69) \\
$E_{\rm x}(1^+_1 1)$       &  0.981      &  0.86 &  1.00 (16;03) \\
$E_{\rm x}(3^+_1 1)$       &  2.255(3) &  2.86 &  2.75 (16;09) \\
$E_{\rm x}(0^+_1 1)$       &                  &  2.51 &  4.01 (84;20) \\
$E_{\rm x}(1^+_2 1)$       &  3.210      &  3.33 &  4.73 (84;21) \\
$E_{\rm x}(2^+_2 1)$       &                  &  3.78 &  4.78 (44;12) \\
$E_{\rm x}(2^+_3 1)$       &                  &  5.22 &  5.94 (37;20) \\
$E_{\rm x}(1^+_3 1)$       &  5.400     &  4.81  &  6.09 (70;22) \\
$E_{\rm x}(4^+_1 1)$       &  6.53(20) &  6.44 &  7.45 (36;15) \\
$E_{\rm x}(3^+_2 1)$       &                  &  7.31 &  8.24 (50;22) \\
$E_{\rm x}(0^+_1 2)$       & 10.822    & 11.25 & 11.77 (27;29) \\
    \end{tabular}
  \end{ruledtabular}
\end{table}

One noticeable difference between
the chiral $NN$ and the chiral $NN+NNN$ predictions appears among the low-lying levels -- an interchange of the order of the $0^+_1$ and $3^+_1$ states.  We note
that the $0^+_1$ state has not been observed experimentally. However, the recent Ref.~\cite{Mitchell10} does claim observation of the low-lying $0^+$ resonance based on the R-matrix analysis of the $p$-$^7$Be scattering experiment performed in the energy range between 1.6 to 2.8 MeV in the center of mass. They suggest the $0^+_1$ resonance is at 1.9 MeV which places it below the experimental $3^+_1$ state.   However, our calculated $0^+_1$ energy obtained with the chiral $NN$ and $NNN$ interaction is above our calculated  $3^+_1$ state. 
On the other hand, note that this $0^+_1$ state has a larger $\hbar\Omega$ dependence than lower-lying states which suggests a proper scattering treatment is needed for its properties.  It is known that the positions of resonances are affected by the coupling to the continuum as demonstrated, e.g. in Ref.~\cite{NCSMRGM_Be7} where the $^8$B and $^8$Li resonances were investigated within the NCSM/RGM approach.

In Table \ref{tab:B8Li8energies} we also predict a significant number of additional levels in these $A=8$ systems that are not yet known experimentally. The ordering of these predicted levels is sensitive to the presence of the NNN interaction so it would be very valuable to have additional experimental information on these states.  We also note that many of these predicted states have larger basis-space dependences which are dominated by their
HO frequency dependence in the $N_{max}=8$ basis space.  This suggests that these continuum states may be somewhat broader resonances than the established states since increasing frequency dependence in HO basis treatments of resonances has been correlated with increasing resonance width \cite{Shirokov09} as mentioned above.

In Table \ref{tab:B8Li8moments} we compare $A=8$ experimental and theoretical results for a selection of electromagnetic observables as we did above for the $A=7$ nuclei.
Here, the radii and quadrupole
moments are somewhat larger and closer to experiment in our
chiral $NN$calculations due, in part, to weaker binding. 
In addition, contrary to the $^7$Be-$^7$Li case, we
observe here an interesting difference between the $NN$ and $NN+NNN$ cases for
the magnetic moment prediction. By including the $NNN$ interaction the
magnetic moment of $^8$Li is significantly greater than that of $^8$B in agreement
with experiment, while the $NN$ interaction alone predicts the opposite.
Clearly, our results suggest that the $A=8$ magnetic moments are sensitive to a presence
of three-nucleon interaction in the Hamiltonian. 
These $A=8$ magnetic moments also show sensitivity to $\hbar\Omega$ since they change
by 10\% - 20\% in the largest basis space over the range of $\hbar\Omega$ values in Table \ref{tab:B8Li8moments}.

Next, consider the
B(M1;$1^+\rightarrow 2^+$) transitions presented in Table \ref{tab:B8Li8moments}.  
Here, the calculated matrix elements are 20\% - 35\% smaller than the experimental
values.  More significantly, the calculated results are nearly unchanged when 
the $NNN$ interaction is included.  Both these features are reminiscent of the 
B(M1;$\frac{1}{2}^-\rightarrow\frac{3}{2}^-$) transitions in the $A=7$ nuclei shown
in Table \ref{tab:Be7moments}.

\begin{table}[hbtp]
  \caption{The $^8$B and $^8$Li point-proton rms radii (in fm), ground-state quadrupole (in $e$fm$^2$)
and magnetic (in $\mu_{\rm N}$) moments and M1 transitions (in $\mu_{\rm N}^2$)
obtained within the NCSM for different HO frequencies (given in MeV)
and model spaces for the chiral $NN$ and chiral $NN+NNN$ interactions. 
Most experimental values
are from Ref. \cite{A=8}. The point proton rms
radius $r_{\rm p}$ for $^8$Li is evaluated from the experimental rms charge radius
 \cite{Nortershauser:2011zz} as discussed in the text. 
  \label{tab:B8Li8moments}}
  \begin{ruledtabular}
    \begin{tabular}{cccccc}
\multicolumn{6}{c}{$^8$B} \\ \hline
\multicolumn{6}{c}{$NN$} \\
$\hbar\Omega$ & $N_{\rm max}$ & $r_{\rm p}$ & Q & $\mu$ & B(M1; $1^+\rightarrow 2^+$) \\
\hline
13 &  6 & 2.409 &  5.243 & 1.416 & 3.175 \\
13 &  8 & 2.463 &  5.569 & 1.438 & 3.202 \\ \hline
\multicolumn{2}{c}{Expt.} &  - - - & 6.83(21) & 1.0355(3) & 4.71(21) \\ \hline
\multicolumn{6}{c}{$NN+NNN$} \\
$\hbar\Omega$ & $N_{\rm max}$ & $r_{\rm p}$ & Q & $\mu$ & B(M1; $1^+\rightarrow 2^+$) \\
\hline
11 &  6 & 2.472 & 5.174 & 1.296 & 3.215 \\
11 &  8 & 2.484 & 5.294 & 1.278 & 3.253 \\ \hline
%
13 &  6 & 2.335 & 4.464 & 1.146 & 3.197 \\
13 &  8 & 2.374 & 4.698 & 1.157 & 3.215 \\ \hline
%
15 &  6 & 2.253 & 4.004 & 1.042 & 3.187 \\
15 &  8 & 2.308 & 4.317 & 1.079 & 3.183 \\ \hline
\multicolumn{2}{c}{Expt.} & - - - & 6.83(21) & 1.0355(3) & 4.71(21) \\
\hline
\multicolumn{6}{c}{$^8$Li} \\ \hline
\multicolumn{6}{c}{$NN$} \\
$\hbar\Omega$ & $N_{\rm max}$ & $r_{\rm p}$ & Q & $\mu$ & B(M1; $1^+\rightarrow 2^+$) \\
\hline
13 &  4 & 2.119 & 2.539 & NA & 4.143 \\
13 &  6 & 2.117 & 2.700 & 1.272 & 4.133 \\
13 &  8 & 2.147 & 2.826 & 1.249 & 4.125 \\ \hline
\multicolumn{2}{c}{Expt.} & 2.23(6) & +3.27(6) & +1.653560(18)  & 5.0(1.6) \\ \hline
\multicolumn{6}{c}{$NN+NNN$} \\
$\hbar\Omega$ & $N_{\rm max}$ & $r_{\rm p}$ & Q & $\mu$ & B(M1; $1^+\rightarrow 2^+$) \\
\hline
11 &  4 & 2.221 & 2.687 & NA & 4.153 \\
11 &  6 & 2.177 & 2.766 & 1.378 & 4.186 \\
11 &  8 & 2.168 & 2.830 & 1.392 & 4.191 \\ \hline
13 &  4 & 2.080 & 2.453 & NA & 4.151 \\
13 &  6 & 2.064 & 2.549 & 1.500 & 4.145 \\
13 &  8 & 2.085 & 2.648 & 1.487 & 4.127 \\ \hline
15 &  4 & 1.984 & 2.245 & NA & 4.172 \\
15 &  6 & 2.001 & 2.394 & 1.581 & 4.111 \\
15 &  8 & 2.041 & 2.529 & 1.543 & 4.072 \\ \hline
\multicolumn{2}{c}{Expt.} & 2.23(6) & +3.27(6) & +1.653560(18)  & 5.0(1.6) \\
    \end{tabular}
  \end{ruledtabular}
\end{table}

\begin{figure}[t]
\includegraphics*[width=1.0\columnwidth]{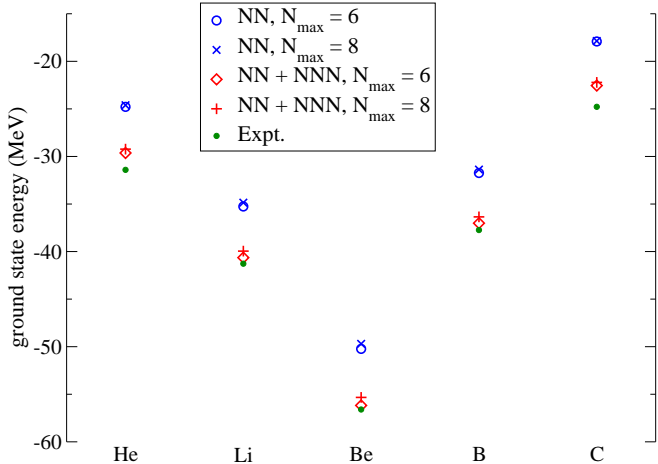}
\caption{(Color online) Calculated and experimental  
  gs energies for $A = 8$.  Calculated results are obtained
  with the chiral EFT $NN$ and $NN$ plus $NNN$ interactions at $\hbar\Omega = 13
  MeV$ and $N_{max} = 6$ and $8$. \label{groundstate}} 
\end{figure}

Our calculated gs energies of $A=8$ nuclei are summarized in
Table~\ref{tab1} and shown in Fig.~\ref{groundstate}.  With both the NN interaction alone
and the NN + NNN interaction, the results appear
reasonably stable in going from $N_{max} = 6$ to $N_{max} = 8$.  
The role of the chiral EFT NNN interaction is to shift all calculated gs
energies significantly closer to the experimental results.  There appears to be a tendency to
underbind these nuclei as one moves away from the minimum in the valley of stability. 

\begin{table}
\begin{tabular}{cccc}
        &  Exp & $NN$ & $NN+NNN$ \\ \hline
$^8$He  & -31.408 & -24.61 & -29.23 (41) \\
$^8$Li    & -41.277 & -34.86 & -39.95 (69) \\ 
$^8$Be  & -56.499 & -49.70 & -55.33 (84) \\
$^8$B    & -37.737 & -31.38 & -36.35 (67) \\
$^8$C    & -24.782 & -17.86 & -22.22 (33) \\
\end{tabular}
\caption{The NCSM results in the $8\hbar\Omega$ basis space for the
  gs energies, in MeV, of $^8$He, $^8$Li, $^8$Be, $^8$B and
  $^8$C using the chiral $NN$ and the chiral $NN+NNN$ interactions. 
  Experimental energies 
  are from Ref ~\cite{energ1}. 
  The basis space dependencies, explained in the text, are in parenthesis. }
\label{tab1}
\end{table}

\begin{table}
  \caption{The $^8$Be ground and excited state energies (in MeV)
obtained using the chiral $NN$ and chiral $NN+NNN$ interactions.
The HO frequency of $\hbar\Omega=13$ MeV and the $8\hbar\Omega$ model space were used.
Experimental values are from Refs. \cite{A=8,energ1}.
States flagged with the superscript ($^a$) have significant isospin mixing 
in both the experimental and theoretical results.
We quote the isospin assigned by Ref.   \cite{A=8}.
Uncertainties, explained in the text and in the caption to Table \ref{tab:Be7energies}, 
are in parenthesis.
  \label{tab:Be8energies}}
  \begin{ruledtabular}
    \begin{tabular}{cccc}
\multicolumn{4}{c}{$^8$Be} \\
            & Expt.   & $NN$ & $NN+NNN$ \\
             \hline
$|E_{\rm gs}(0^+ 1)|$       & 56.499 & 49.70 & 55.33 (84) \\
$E_{\rm x}(2^+ 0)$           &  3.03  &  3.45   &  3.63 (10;10) \\
$E_{\rm x}(4^+ 0)$           & 11.35 & 11.87 & 12.53 (30;33) \\
$E_{\rm x}(2^+ 0+1)$$^a$   & 16.63 & 16.30 & 17.01  (13;18) \\
$E_{\rm x}(2^+ 0+1)$$^a$   & 16.92 & 16.54 &  16.77 (12;23) \\
$E_{\rm x}(1^+ 1)$$^a$   & 17.64 & 17.23 & 17.91 (07;21) \\
$E_{\rm x}(1^+ 0)$$^a$   & 18.15 & 16.87 & 18.25 (27;24) \\
$E_{\rm x}(3^+ (1))$         & 19.07 & 19.20 & 19.67 (29;25) \\ 
$E_{\rm x}(3^+ (0))$         & 19.24 & 18.59 & 20.07 (25;21) \\ 
$E_{\rm x}(0^+ 1)$           &             & 18.78 & 20.84 (72;39) \\
$E_{\rm x}(4^+ 0)$           & 19.86 &               & 20.47 (50;33) \\
    \end{tabular}
  \end{ruledtabular}
\end{table}

\begin{figure}[t]
\includegraphics*[width=1.0\columnwidth]{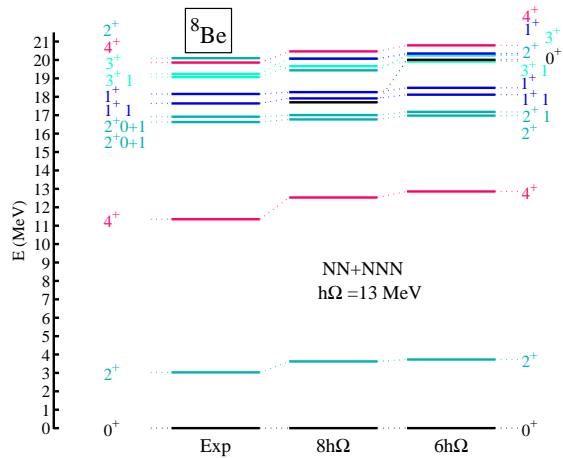}
\caption{(Color online) Calculated and experimental excitation energies of $^{8}$Be. Dependence on the size of the basis is presented. The chiral EFT $NN$ and $NNN$ interaction was used. The isospin of the states is $T=0$ unless shown otherwise. See the text for further details.}\label{fig:Be8_spectrum}
\end{figure}

Our calculated excitation levels of $^8$Be are compared to experiment
in Table~\ref{tab:Be8energies} and Fig.~\ref{fig:Be8_spectrum}. We
note a good agreement of the level ordering compared to experiment and
also a good stability of the spectrum with respect to the change of
the model space size. An exception in this regard is the calculated
first excited $0^+ 0$ state that is an intruder state with large
multi-$\hbar\Omega$ components. The appearance of this state and the
corresponding $2^+$ and $4^+$ excitations were discussed in detail in
Ref.~\cite{Be8_intruder}. We also note that the existence of a $2^+$
intruder broad resonance was confirmed in the R-matrix analysis of
the reactions with $^8$Be as the composite system~\cite{Rmatrix}.

We note that our present results for $^7$Li, $^7$Be, $^8$Li and $^8$B compare favorably with those of Ref.~\cite{Be7p} obtained with the CD-Bonn 2000~\cite{cdb2k} and the INOY~\cite{INOY} $NN$ potentials. 
For example, the trend for the magnetic moment in $^{8}$B to decrease when one adds the
$NNN$ interaction is similar to the decrease obtained as one changes from the CD-Bonn to
the INOY interaction.  The magnitudes are also comparable.
Earlier work \cite{Bogner:2007rx} with the chiral NN interaction alone \cite{Epelbaum,Entem:2003ft,Machleidt:2011zz} investigated the gs energy of $^7$Li as a function of the scale set in the Similarity Renormalization Group
(SRG) approach to the effective NN interaction  and found the gs energy to range from -37.8 MeV to -42.0 MeV which spans our own result of -38.60(44) MeV in Table \ref{tab:Be7energies}.  The dependence on renormalization scale implies that higher-body effective interactions are required to obtain a stable result. 

Extensive {\it ab initio} calculations of A=7 and A=8 nuclei have been performed with Variational Monte-Carlo and Green's Function Monte Carlo~\cite{Pieper:2004qw,Pervin:2007} methods using NN plus NNN interactions derived from meson-exchange theory.  These works provide the most precise agreement between theory and experiment for the observables they investigate and are closer to convergence, in our view.  In addition, they have evaluated meson-exchange current corrections to electroweak processes. Though our agreement with experiment is not as good and we do not yet incorporate exchange current corrections, we are able to expand the suite of observables to compare with experiment and to provide the platform for systematic improvements with anticipated future developments of the chiral interactions and exchange current corrections in the chiral approach.

More recently, a comprehensive review of the Unitary Correlation Operator Method (UCOM) 
\cite{Roth:2010bm} presents extensive results for light nuclei including $^7$Li using the Argonne V18 NN interaction \cite{Wiringa:1994wb}.  With the UCOM method they obtain a variational upper bound on the $^7$Li ground state of -37.4(6) MeV which is also close to our result though there are many differences in the calculations. Their spectra for $^7$Li through the first seven excited states are also in the experimental order but spread more than experiment as are our own spectra.

Among other efforts with {\it ab initio} no-core methods to address some of the same nuclei investigated here with, we mention our efforts with an NN interaction derived by inverse scattering methods, JISP16 \cite{JISP16} applied to $^7$Li and $^8$Li \cite{Cockrell} using the no-core full configuration (NCFC) method \cite{NCFC}.  Results of those investigations also appear to be in rough accord with the results presented here.  For example, the gs energy of $^7$Li and $^8$Li are -38.253(1) MeV and -39.485(16) MeV respectively, both within 500 keV of the results we report here.

\section{Conclusions}
\label{conclusions}

In this work, we used the Okubo-Lee-Suzuki renormalization of the chiral Hamiltonian specific to each model space employed and presented results for $A=7$ and $A=8$ nuclei. Our results demonstrate that the $NNN$ interaction improves the agreement with experimental data not only for binding energies but also for excitation energies and other observables. Among other features, our results suggest that the $A=8$ magnetic moments are especially sensitive to the presence of three-nucleon interaction in the Hamiltonian. 

Taking into account our estimates of the basis-space dependence of our spectra given in Tables~\ref{tab:Be7energies} and 
~\ref{tab:B8Li8energies}, we find that there are residual differences between theory and experiment that can now be attributed to the need for further improvements to our approach.  Those improvements could originate from improved chiral 3-body interactions, adding chiral 4-body interactions and/or including effective 4-body interactions.  We also recall that there is imperfect knowledge of the nonperturbative coupling constants 
in the currently-employed chiral NN + NNN interactions that could, if exploited, remove differences between the current theoretical results and experiment.

The present results will be useful for comparing with calculations performed with the SRG-evolved chiral interactions that are currently under way~\cite{SRG_pshell}. In this regard, the extension of the model space to the $8\hbar\Omega$ basis is significant as already proven in the case of $^6$Li calculations~\cite{Li6_SRG}. 

We have now extended the {\it ab initio} no-core shell model calculations with two- and three-nucleon forces in the complete $8\hbar\Omega$ basis to all $p$-shell nuclei by technical advances of the shell model code MFDn and the codes that calculate and transform the three-body interaction matrix elements. As demonstrated in the  $8\hbar\Omega$ results for the $A = 14$ nuclei ~\cite{C14} we have now the capability to calculate any $p$-shell nucleus in model spaces up to $8\hbar\Omega$ with matrix dimensions exceeding 1 billion with Hamiltonians that include $NNN$ interactions.  This basis extension capability is also significant for a further refinement of the importance-truncation approach~\cite{Roth09} and for the nuclear reaction applications within the NCSM/RGM method~\cite{NCSMRGM}. 
 
We also note that further improvements in the three-body interaction transformation algorithm and a new division of tasks between the shell-model and the three-body interaction transformation codes will allow one to reach even higher $N_{\rm max}\hbar\Omega$ basis spaces.  This is significant since both the importance truncation approach~\cite{Roth09,Roth} and the SU3-NCSM~\cite{Dytrych} provide access to much higher $N_{\rm max}\hbar\Omega$ basis spaces with chiral EFT interactions. \\

\acknowledgments
This work was supported in part by US DOE Grants DE-FC02-09ER41582 (UNEDF SciDAC Collaboration), DE-FG02-87ER40371 and US NSF grant 0904782. This work was also repared in part by LLNL under Contract DE-AC52-07NA27344.
Computational resources were provided by the Lawrence Livermore National Laboratory (LLNL) Institutional Computing Grand Challenge program and
by the National Energy Research Scientific Computing Center (NERSC), which is supported by the Office of Science of the U.S. Department of Energy under Contract No. DE-AC02-05CH11231.

\end{document}